\documentclass[aps,prapplied,reprint,superscriptaddress,twocolumn]{revtex4-2}
\usepackage[utf8]{inputenc}
\usepackage[T1]{fontenc}
\usepackage{lmodern}
\usepackage{textcomp, mathtools, amssymb, amsthm, wasysym, gensymb}
\usepackage[colorlinks=true, linkcolor=red, citecolor=blue, urlcolor=blue]{hyperref}

\newcommand\unine{Laboratoire Temps-Fréquence, Institut de Physique, Université de Neuchâtel, Avenue de Bellevaux 51, 2000 Neuchâtel, Switzerland}
\newcommand\ethz{Institute for Quantum Electronics, ETH Zurich, Auguste-Piccard-Hof 1, 8093 Zurich, Switzerland}

\begin{document}	

\title{Coherent Control of Mid-Infrared Frequency Comb by Optical Injection of Near-Infrared Light}

\author{Kenichi~N.~Komagata}
\email[Corresponding author: ]{kenichi.komagata@unine.ch}
\affiliation{\unine}

\author{Alexandre~Parriaux}
\affiliation{\unine}

\author{Mathieu~Bertrand}
\affiliation{\ethz}

\author{Johannes~Hillbrand}
\affiliation{\ethz}

\author{Mattias~Beck}
\affiliation{\ethz}

\author{Valentin~J.~Wittwer}
\affiliation{\unine}

\author{Jérôme~Faist}
\affiliation{\ethz}

\author{Thomas~Südmeyer}
\affiliation{\unine}

\date{\today}

\begin{abstract}
We demonstrate the use of a low power near-infrared laser illuminating the front facet of a quantum cascade laser (QCL) as an optical actuator for the coherent control of a mid-infrared frequency comb. We show that by appropriate current control of the QCL comb and intensity modulation of the near-infrared laser, a tight phase lock of a comb line to a distributed feedback laser is possible with 2~MHz of locking bandwidth and 200~mrad of residual phase noise. A characterization of the whole scheme is provided showing the limits of the electrical actuation which we bypassed using the optical actuation. 
Both comb degrees of freedom can be locked by performing electrical injection locking of the repetition rate in parallel. However, we show that the QCL acts as a fast near-infrared light detector such that injection locking can also be achieved through modulation of the near-infrared light. These results on the coherent control of a quantum cascade laser frequency comb are particularly interesting for coherent averaging in dual-comb spectroscopy and for mid-infrared frequency comb applications requiring high spectral purity.
\end{abstract}

\maketitle

\section{Introduction}
Quantum cascade lasers (QCL) emitting frequency combs were first demonstrated in 2012~\cite{Hugi-nature-2012} and have since then  become an established technology for fast and broadband mid-infrared (MIR) spectroscopic applications, such as time-resolved studies of microsecond-scale molecular dynamics~\cite{klocke2018singleshot}, high-pressure and temperature thermometry in shock tubes~\cite{pinkowski2021thermometry}, or high-resolution measurements of molecular spectra~\cite{lepere2022midinfrared,agner2022highresolution,komagata2023absolute}. Some of their key assets are their low footprint and large optical power compared to other types of MIR combs, especially in the 8--10~µm spectral range that is usually reached via nonlinear frequency conversion~\cite{iwakuni2018phasestabilized,krzempek2019stabilized,Hoghooghi-lsa-2022}. Indeed, QCLs are electrically-driven devices that directly emit frequency combs in the MIR with a power that can reach 1~W~\cite{jouy2017dual}.

In the context of comb spectroscopy, the use of two mutually coherent combs with slightly different repetition rates, namely dual-comb spectroscopy (DCS), has shown great potential for ultra fast and high resolution measurements without the need of complex and expensive instruments~\cite{Coddington-optica-2016,Picque-natphot-2019}. DCS has been well demonstrated with mode-locked lasers in the near-infrared (NIR), but the MIR is more interesting as molecules generally have stronger absorption features, which is advantageous for many applications such as trace gas detection~\cite{Giorgetta-LRP-2021}, or isotope ratio measurements~\cite{Parriaux-prr-2022}. 
DCS with QCLs is then highly interesting as it provides a compact, low footprint and high-resolution spectrometer~\cite{lepere2022midinfrared,agner2022highresolution,komagata2023absolute}.

In DCS, Comb stabilization is not strictly necessary thanks to the availability of computational phase correction of free-running lasers~\cite{sterczewski2019computational}. Nevertheless, stabilization of the combs allows accurate frequency referencing and arguably more flexibility, e.g., smaller repetition rates. Moreover, to properly establish QCLs as a source of choice for MIR comb applications such as optical frequency synthesis~\cite{jost2002continuously,argence2015quantum,komagata2022absolute}, demonstrating high spectral purity and coherent control can be considered as important as increasing their bandwidth and detecting their offset frequency~\cite{telle1999carrierenvelope}.

For QCLs, actuation on the drive-current is the most straightforward way to phase-lock a comb line~\cite{cappelli2016frequency,westberg2017midinfrared,Komagata-oe-2021}, while the other degree of freedom, namely the repetition rate, is locked by electrically injecting a radio-frequency (RF) signal close to the round-trip frequency~\cite{Hillbrand-natphot-2019}. These two handles allow the coherent control of QCL combs and can be used together~\cite{komagata2022absolute}. However, as for distributed feedback (DFB) QCLs~\cite{tombez2012temperature,tombez2013wavelength}, we expect the intrinsic frequency noise of the QCL comb and the stabilization by drive current to involve the same physical process, i.e., current to temperature to refractive index change. In that case, the time scale of the stabilization will always be close to the cutoff of the noise process, thus limiting the performance of the lock. A solution adopted for other types of lasers is to employ another actuator bound by time scales much faster than the noise source such as opto-optical modulation~\cite{hoffmann2013ceo}.

For QCLs, light illumination at another wavelength also enables its control and can play as an actuator. The technique has been used on DFB-QCLs emitting a single wavelength for applications such as fast switching~\cite{zervos2007alloptical}, gain enhancement~\cite{chen2009quantum}, stabilization~\cite{Tombez-ol-2013}, and frequency modulation~\cite{Peng-ptl-2016}. More recently, a QCL comb emitting in the THz range was locked via the intensity modulation of a white LED~\cite{Consolino-lpr-2021}. We also note that illumination of resonant light enabled the mutual lock between two MIR QCL combs via injection locking~\cite{Hillbrand-oe-2022}. However, this only allows mutual locking at the same optical frequency, whereas modulation by off-resonant light offers more possibilities. 

Indeed, light illumination at a different wavelength can influence various parameters with different strengths, for example to allow pure frequency modulation of a DFB-QCL~\cite{peng2017purified}. Moreover, multiple locking-scheme could be possible for combs, such as the locking of the repetition rate via current actuation~\cite{cappelli2019retrieval,shehzad2020frequency} combined with locking of the comb line frequency with NIR light. 
Finally, the NIR is supported by mature technologies allowing a wide range of possibilities. Among others, the high modulation bandwidths reaching a few tens of GHz could enable the injection locking of the QCL comb repetition frequency.
In light of the above, there is a compelling interest to investigate the potential of NIR light illumination for the coherent control of MIR QCL combs.

In this work, we characterize the influence of NIR light illuminating the front facet of a MIR QCL comb emitting in the 8~\textmu{}m range. We measure its transfer function on the comb frequencies and compare it to the more conventional electrical actuation. We demonstrate that the limits of electrical actuation for phase-locks can be bypassed using the intensity modulated NIR light by tightly locking a QCL comb to a DFB-QCL. Lastly, we show that the repetition rate of the QCL can also be injection-locked by intensity modulation of the NIR light.

\section{Experimental setup}

\begin{figure}
\centering
\includegraphics[width=\linewidth]{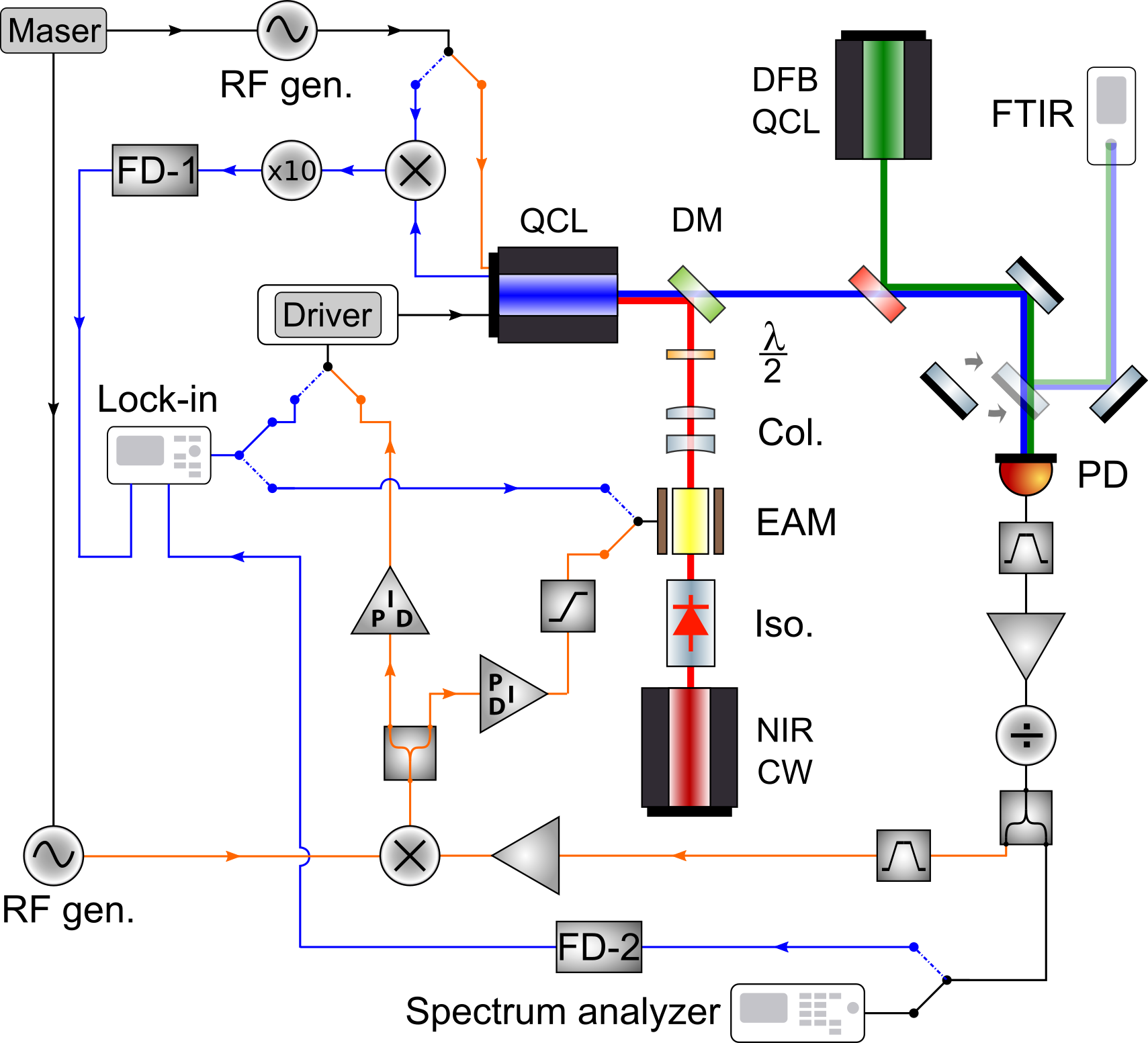}
\caption{Schematic showing the experimental setup used to phase lock a line from a QCL comb to a DFB-QCL using an intensity modulated NIR CW laser, and for characterization. The electrical blue wires represent the path for characterizing the repetition frequency and a comb line, whereas the orange wires represent the path for phase locking. CW: continuous wave, EAM: electro-absorption modulator, DM: dichroic mirror, FD: Frequency discriminator, PD: Photodetector, FTIR: Fourier transform infrared spectrometer.}
\label{fig:setup}
\end{figure}

The experimental setup considered here is presented in \autoref{fig:setup} and pivots around a QCL comb.
The laser is controlled in current and temperature with a custom-made driver that sets the operation point of the laser to 1200~mA (1.45 times the lasing threshold) and 0\celsius{}. A frequency comb centered around 1305~cm$^{-1}$ is emitted, with approximately 80 lines, a total power of 126~mW, and a repetition frequency of 11.057~GHz. In QCL combs, the comb modes beating together in the Fabry-Perot cavity lead to a measurable voltage oscillating at the repetition frequency~\cite{piccardo2018timedependenta}. Thus, two wire bonds connect the top of the QCL waveguide near the front and back facet of the laser to RF waveguides on a PCB chip to efficiently inject and extract the repetition rate independently of the drive current.

A custom-made dichroic mirror produced by ion beam sputtering is placed in the optical comb path to direct NIR light from a continuous wave (CW) laser at 1.55~µm (Optilab, DFB-1550-EAM-12-K) to the front facet of the QCL via the collimation lens (Thorlabs, C037TME-F). The NIR laser has an output power around 1~mW and can be modulated up to 12~GHz using an integrated electro-absorption modulator (EAM). 
The beam is aligned into the QCL so as to maximize the frequency response (see Sect.~\ref{sect:response}). The amount of light effectively reaching the front facet is 55\% of the emitted power of the NIR laser and these losses are mainly due to the transmission of the QCL collimation lens at 1.55~µm. The polarization of the NIR light was fixed, however, it did not seem to change the experimental results.

\begin{figure}
\centering
\includegraphics[width=\linewidth]{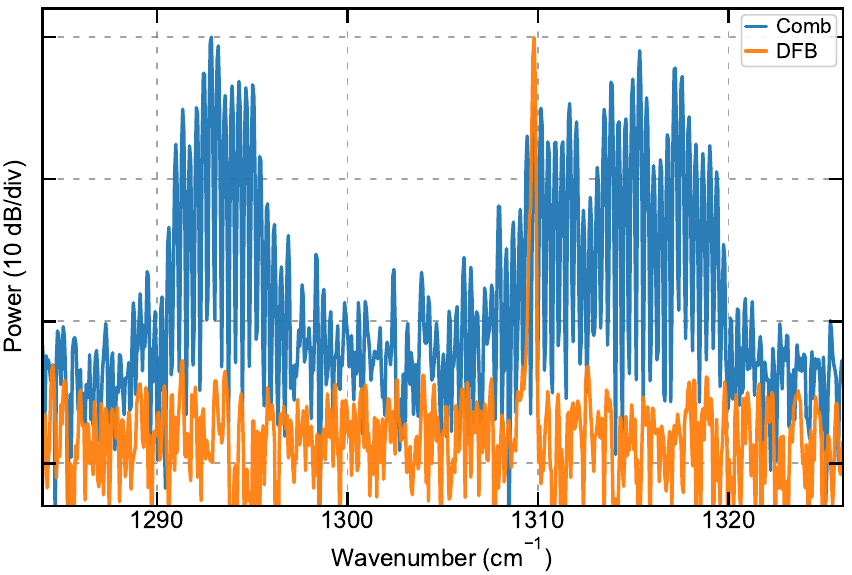}
\caption{Typical optical spectrum generated by the QCL comb and the continuous wave generated by the DFB-QCL.}
\label{fig:opt_spectrum}
\end{figure}

After passing through the dichroic mirror, the comb is mixed on a 50/50 beam splitter with a CW MIR light generated by a DFB-QCL (Alpes Laser). The latter is driven at a current of 191~mA and a temperature of 0~\celsius{}, to emit at a frequency $f_\text{cw} = $ 1309.79~cm$^{-1}$, which is within the spectral range of the frequency comb. 
A typical optical spectrum of the comb and the DFB recorded with a Fourier transform infrared spectrometer (Bristol 771A-MIR) is presented in \autoref{fig:opt_spectrum}.
The beating $f_b$ between the comb and the DFB-QCL is recorded on a fast photodetector (VIGO, PV-4TE-10.6).

In the following section we will start with studying the frequency response of the QCL comb when NIR light illuminates its front facet
\section{Response characterization}\label{sect:response}
We start with the static response before moving on to the frequency dependent response.

\subsection{Static response}\label{sect:static_response}
\begin{figure}[b]
\centering
\includegraphics[width=\linewidth]{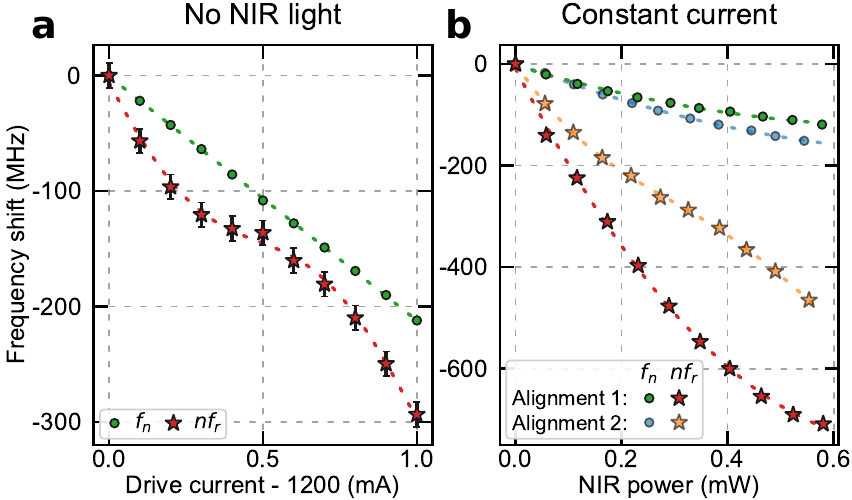}
\caption{Static response of the QCL comb frequencies to a change in (a) drive current or (b) illuminated NIR power. The shift in $f_r$ is scaled by $n$. In (b), the response is reported for two alignment conditions (see main text). Dashed lines are fitted functions to the experimental data. Error bars are plotted when larger than the data marker.}
\label{fig:static_response}
\end{figure}

First, we slowly vary the NIR power reaching the QCL from 0~mW to 0.6~mW, and measure on a RF spectrum and phase noise analyzer (Rohde \& Schwarz, FSWP26) the frequency shift of a comb line $f_n$ via its beating with the DFB-QCL, and of the repetition rate $f_r$, measured directly via the independent channel for RF extraction on the QCL comb. For comparison, we also measure the frequency shifts when the drive current of the QCL comb is varied over 1~mA. These results are presented in \autoref{fig:static_response}(a, b), where the shift in $f_r$ is scaled by the mode number $n=3550$ for better comparability with $f_n$.

For a small variations (1~mA) of the QCL drive currents, the frequencies shift linearly with a superimposed sinusoidal modulation on $f_r$ due to back-reflections~\cite{Teng-ol-2023}. The fitted functions to the experimental data (dashed lines in \autoref{fig:static_response}) using a linear model supplemented by a sine wave give an average slope of -210~MHz/mA and -85~kHz/mA for $f_n$ and $f_r$ respectively, although the slope for $f_r$ is dependent on the drive current (position within the modulation) due to the back reflections. The response to low power NIR light suggests a quadratic trend. Moreover, the shift ratio between $f_n$ and $nf_r$ depends on the alignment of the NIR light on the active region of the QCL comb. The main alignment method used in this article (alignment~1, described in \autoref{sect:dynamic_response}) is to maximize the dynamic response of $f_n$ at modulation frequencies of 100~kHz. Another method would be to maximize the static shift of $f_n$ (alignment~2). In the latter case, $f_r$ shows a sinusoidal modulation that is nearly nonexistent for alignment~1, and varies far less than for alignment~1. For alignment~1, the slope interpolated at zero power are -315~MHz/mW and -570~kHz/mW for $f_n$ and $f_r$ respectively.

\subsection{Dynamic response}\label{sect:dynamic_response}
We now take an interest in the frequency-dependent response of the QCL comb. A lock-in amplifier (LIA, Zurich Instruments, UHFLI) modulates the power of the NIR light via the EAM or the comb drive current through the laser driver, see the blue path in \autoref{fig:setup}. We study the response of three comb characteristic frequencies, namely, the offset frequency $f_0$, $f_r$, and $f_n$. 
Naturally, the three frequencies are coupled to each other through the comb equation:
\begin{equation} \label{eq:comb}
    f_n = f_0 + n f_r \quad ,
\end{equation}
where $n$ is an integer. 
Due to the modulation of the drive current or the NIR power set by the LIA at frequency $\omega$, the comb, i.e., its frequencies respond as: 
\begin{equation} \label{eq:mod}
    f_i = f_i^{(0)} + \Delta f_i \sin{(\omega t + \theta_i)} \quad ,
\end{equation}
where $i=\{0 ; r ; n\}$ indexes the three comb frequencies under study,  $f_i^{(0)}$ is the average value of $f_i$, $\Delta f_i$ the peak amplitude of $f_i$ due to the modulation, and $\theta_i$ is the phase of the response. To measure the amplitudes and phases, we convert the frequency modulation of $f_i$ to a voltage modulation using a frequency discriminator (FD) and demodulate this voltage on the LIA. 

For this purpose, the repetition rate of the comb is extracted electrically as before, amplified and down-mixed to 60~MHz. As $\Delta f_r$ is only on the order of 20~kHz, we take the 10\textsuperscript{th} harmonic and down-mix it to 21~MHz before sending it to FD-1 (Miteq, FMDM-21.4/4-2, see \autoref{fig:setup}), whose output is connected to the LIA. 
As for the comb line $n$, the amplitude and phase response of the comb line $f_n$ is encoded in the beating signal with the DFB-QCL as:  
\begin{equation}
f_b = f_n^{(0)} +  \Delta f_n \sin{(\omega t + \theta_n)}  - f_\text{cw} \quad .
\end{equation}
The signal $f_b$ detected on the photodetector is then filtered, amplified, divided by 15 and 3 (RF bay, FPS-15-8, FPS-3-8), up-mixed to 60~MHz (not shown), and fed to a FD (Miteq, FMDM-60/16-4BC), whose output is connected to the LIA, as shown in \autoref{fig:setup}. The division step allows a larger frequency excursion to be measured than the linear range of the FD. The frequency to voltage conversion ratios, including all division and multiplication steps for $\Delta f_r$ and $\Delta f_n$ are respectively ($8.2\pm 0.4$)~V/MHz and ($5.93\pm 0.06$)~mV/MHz.

Regarding $f_0$, this frequency can not be directly detected in practice since $f$--to--$2f$ interferometry~\cite{telle1999carrierenvelope} is currently unavailable to QCL combs but fluctuations thereof can be detected~\cite{shehzad2020frequency}, which allows the measurement of its transfer function. However, for experimental simplicity, we can compute it from the transfer function of $f_n$ and $f_r$. Indeed, according to Eq.~\eqref{eq:comb},~\eqref{eq:mod} and the elastic tape model~\cite{telle2002kerrlens,dolgovskiy2012crossinfluence}, we have:
\begin{equation} \label{eq:elastic_tape}
    \begin{aligned} 
   \Delta f_0 \sin{(\omega t + \theta_0)} =&  \Delta f_n \sin{(\omega t + \theta_n)} \\
   & - n \Delta f_r \sin{(\omega t + \theta_r)}    
\end{aligned} \quad .
\end{equation}
Therefore, Eq.~\eqref{eq:elastic_tape} yields that $\Delta f_0$ and $\theta_0$ are respectively the modulus and phase of the complex value $ \Delta f_n \exp{(i\theta_n)}  -n \Delta f_r \exp{(i\theta_r)} $. Here, $n=3551$ is set by the wavenumber of the DFB-QCL at 1309.79~cm$^{-1}$, and by the repetition rate.

\begin{figure}[hbt]
\centering
\includegraphics[width=\linewidth]{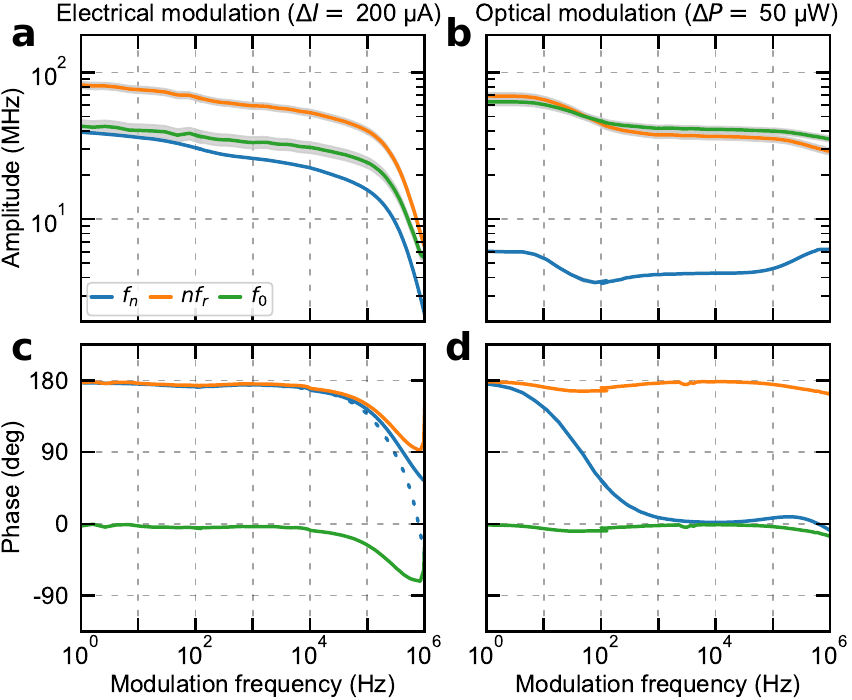}
\caption{Response of the comb frequencies ($f_n, nf_r, f_0$) to modulation of the drive current (a, c) and of the intensity of the NIR light (b, d). Panels (a, b) present the frequency excursions $\Delta f_i$ while panels (c, d) show the phase response $\theta_i$. The dashed line in (c) is the phase response of $f_n$ with the laser driver.}
\label{fig:freq_response}
\end{figure}

\autoref{fig:freq_response} shows the resulting phase $\theta_i$ and amplitude responses $\Delta f_i$ of the frequencies ($f_n, nf_r, f_0$) to modulation of the electrical drive current, $\Delta I$ = 200~\textmu{}A, and the NIR power, $\Delta P$ = 50~\textmu{}W, where $\Delta$ represents the peak amplitude of the modulation. Note that $\Delta f_r$ is scaled by $n$ for better comparison with $\Delta f_n$. Also, the contributions of various components of the characterization scheme (i.e. the FDs and the laser driver) were measured and deducted in order to obtain the laser response as faithfully as possible. Moreover, the responses were measured with two different settings for the ranges [1~Hz, 100~Hz] and [100~Hz, 10~MHz], leading to negligible mismatches at 100~Hz.

Focusing first on the electrical actuation in \autoref{fig:freq_response}(a, c), we observe that the amplitude response $\Delta f_n$ decreases steadily with the modulation frequency setting the 3-dB modulation bandwidth to 30~kHz, before decreasing sharply after about 200~kHz. The phase response remains flat up to 10~kHz with a 90\degree{} modulation bandwidth at 420~kHz, which is coherent with previous tight-locking results~\cite{cappelli2016frequency}. Moreover, the response of $f_n$ closely mimic that of DFB-QCLs~\cite{tombez2013wavelength,hangauer2014high}. $f_r$ follows a similar behavior as $f_n$ with a 3-dB and 90\degree{} modulation bandwidth of 80~kHz and 840~kHz respectively. The dashed line in \autoref{fig:freq_response}(c) is the phase response of $f_n$ with the laser driver, and has an 90\degree{} modulation bandwidth of 280~kHz. The difference between $\Delta f_n$ and $n\Delta f_r$ gives the response $f_0$, which has a flatter response in amplitude with a 3-dB modulation bandwidth of 160~kHz. The uncertainty on the measurements of $\Delta f_n$ and $n\Delta f_r$, in particular, the slope of the FDs which depend on the input RF power induces a large absolute uncertainty on $\Delta f_0$. As for the phase, its response is also flatter, reaching $-70$\degree{} near 1~MHz, after which the measurement is no longer accurate due to the lack of sensitivity. The (quasi-)fix point~\cite{telle2002kerrlens,dolgovskiy2012crossinfluence} increases with modulation frequency from mode number 1850 at 1~Hz to mode number 2200 at 100~kHz.

Although the transfer function of a MIR QCL frequency comb has already been reported in Ref.~\cite{shehzad2020frequency}, our measurements highlight a 90\degree{} modulation bandwidth one order of magnitude above what was previously shown. This is in agreement with the response of DFB-QCLs~\cite{borri2011frequencynoise,tombez2013wavelength,hangauer2014high} and other QCL combs, measured directly~\cite{corrias2022analog} or demonstrated in a phase-lock loop~\cite{cappelli2016frequency,Komagata-oe-2021}.  This discrepancy with the measured modulation bandwidth could be attributed to the frequency response of the bias-tee used in Ref.~\cite{shehzad2020frequency}. Furthermore, we observed in \autoref{fig:static_response} that modulations of $f_r$ due to back reflections locally change the slope, such that the ratio between $\Delta f_0$ and $\Delta f_n$ and the fix points are expected to be modulated as well.

We now turn to the optical actuation shown in \autoref{fig:freq_response}(b, d). For this measurement, the NIR laser was aligned on the QCL to the maximize the dynamic response of $f_n$ at a 100~kHz modulation. We observe a nearly flat phase response for $f_r$ up to 1~MHz, with a small resonance near 33~Hz and a start of a roll-off near 1~MHz. At the resonance near 33~Hz the amplitude response decreases by a factor 2, and then remains nearly flat apart from the onset of the roll-off near 1~MHz. The 33-Hz resonance also marks a change of regime for $f_n$, due to the crossing of $n\Delta f_r$ with $\Delta f_0$, the latter having a similar response as $f_r$, although a smaller relative change in amplitude response, and an opposite phase response.
Thus, for modulation frequencies lower (higher) than 33~Hz, $\Delta f_0$ is smaller (higher) than $n \Delta f_r$, such that the response of $f_n$ flips sign from 180\degree{} to 0\degree{} from 1~Hz to 1~kHz. The respective (quasi-)fixed points are at mode numbers 3250 and 3950. An associated dip in the amplitude response of $f_n$ is also visible. Above 1~kHz, the response is flat up to about 50~kHz, where the response increases in phase and magnitude before showing a signature of a roll-off near 1~MHz.

The mechanism causing frequency modulation due to the drive current change is, identically to DFB-QCLs, a change of the refractive index~\cite{tombez2013wavelength}. In the case of a NIR injection, the strong inter-band absorption in the InGaAs quantum wells (the absorption coefficient $\alpha = 6000$ cm$^{-1}$ at 1550 nm), that are part of the active region of the QCL, is responsible for the modulation of the refractive index through the generation of carriers in the vicinity of the NIR injection point.
Also, we believe that the response below 30~Hz is dominated by thermal processes (heating of the QCL), as the sign of the responses are equal to drive current modulation and as the optimization of maximum static shift of $f_n$ causes modulation of $f_r$ as with drive current changes. 

From the perspective of locking the QCL comb, the actuation via the NIR power is advantageous as it offers a higher bandwidth than the drive current actuation for all comb frequencies, even when the driver response has been accounted for. It is especially well suited for the stabilization of $f_0$ if it can be measured. The stabilization of $f_n$ will require a countermeasure against the sign reversal as detailed in the next section. As for $f_r$, the increased actuation bandwidth is not as interesting given that it can be tightly-locked by actuation on the drive current~\cite{shehzad2020frequency}.

\section{Mutual stabilization}\label{sect:stabilization}
We now seek to implement a mutual lock between one comb line $f_n$ of the QCL comb and a DFB-QCL as a proof-of-principle demonstration. The electrical part of the setup is adapted according to the orange path in \autoref{fig:setup}. The phase fluctuations between the two lasers are obtained by mixing the signal $f_b$ after division by 15 with a synthesized reference frequency locked to the maser. This error signal is injected into two PID controllers (Vescent, D2-125). The output of one PID controller is fed back to the driver of the QCL comb to modulate the electrical current. The output of the second PID controller is fed back, after high-pass filtering with a 2\textsuperscript{nd}-order custom-designed filter with a cutoff at 1~kHz, to the EAM to modulate the intensity of the NIR CW laser. We also monitor the beatnote with the RF spectrum and phase noise analyzer. In this way, slow ($<100$~kHz) corrections of $f_n$, which cannot be handled by the NIR light due to the sign reversal, are done by the drive current, while the NIR light extends the available bandwidth and cancels added noise from the drive current, i.e. the servo bump.

\begin{figure}
\centering
\includegraphics[width=\linewidth]{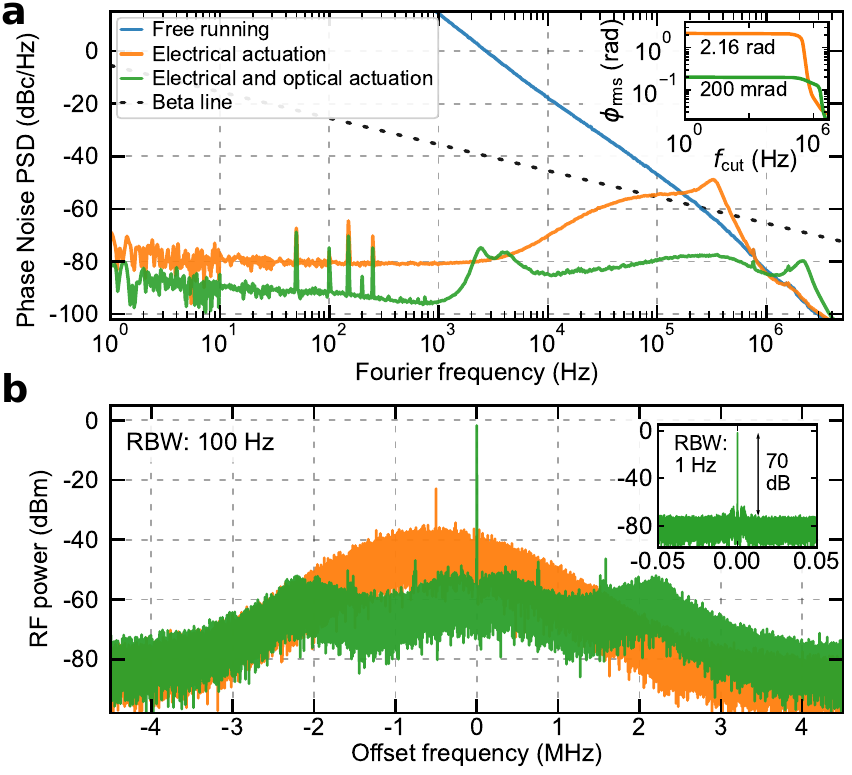}
\caption{Tight-locking of a QCL comb to a DFB-QCL. (a) Phase noise power spectral density (PSD) of the mutual beatnote in free-running mode, electrical actuation, and both electrical and optical actuation. (b) RF power spectrum at 100~Hz RBW of the beatnote between one line of the QCL comb and the DFB-QCL depending on the stabilization scheme used. Inset: Zoom on the beatnote at 1~Hz RBW.}
\label{fig:stabilization}
\end{figure}

\autoref{fig:stabilization}~(a) and (b) respectively show the results of the stabilization of $f_b$ in terms of phase noise and RF power spectrum. When only the electrical actuation is considered, we observe two bumps in the phase noise power spectral density (PSD). The first is due to the limit of the integrator at about 100~kHz, while the second near 300~kHz is the servo bump, which nearly coincides with the 90\degree{} modulation bandwidth of the combined driver and laser (see dashed blue line in \autoref{fig:freq_response}(c)). The white phase noise from 1~Hz to about 100~Hz is due to the reference oscillator, which due to the division by 15, has its contribution increased by 23~dB.
The resulting RF spectrum shows a Gaussian shape topped with a coherent peak.

The addition of the optical actuation reduces the phase noise PSD to a value in the order of -80~dBc/Hz, below the $\beta$-line~\cite{didomenico2010simple}, for all Fourier frequencies. As a result, the integrated phase noise is decreased from 2.16~rad to 200~mrad (see inset in \autoref{fig:stabilization}(a)). The electrical servo bump is eliminated, while a new servo bump appears at 2~MHz. This fast bandwidth allows to employ a lower division ratio of 4: the residual phase noise from 1~Hz to 3~kHz is set by the reference oscillator. At higher frequencies, we believe that the mismatch between the reference voltages of the two different servo controller added noise to the system. Moreover, the bumps near 3~kHz coincide with the cutoff frequency of the high-pass filter.  Further optimization of the electronic components could improve the lock and decrease the integrated phase noise further, including the use of a single dual-output servo controller. We also believe the stabilization bandwidth could be increased by shortening all the cables, fibers, and free-space paths.
In terms of spectrum (\autoref{fig:stabilization}(b)), the power in the coherent peak improves by 20~dB. The height of the pedestal is decreased by 15~dB, such that the difference between the coherent peak and the top of the pedestal is 50~dB at a RBW of 100~Hz, or about 25~dB more than in Ref.~\cite{cappelli2016frequency} at 500~Hz RBW. The inset shows a zoom over the peak at a 1~Hz resolution bandwidth (RBW) with an SNR of 70~dB.

In parallel, the repetition frequency of the comb is stabilized by RF injection locking using a resonant RF signal with 15~dBm of power delivered by a signal generator (Rohde \& Schwarz, SMF100A) referenced to a maser. This RF signal is sent to the QCL via the dedicated channel for RF injection. Thus both degrees of freedom are tightly-locked simultaneously with low residual phase noise. 
The electrical injection of the repetition rate is the usual approach for its stabilization. However, the previous results suggest that the repetition rate could also be injection-locked by modulating the NIR light.

\section{Repetition rate injection via NIR illumination}
\begin{figure}
\centering
\includegraphics[width=\linewidth]{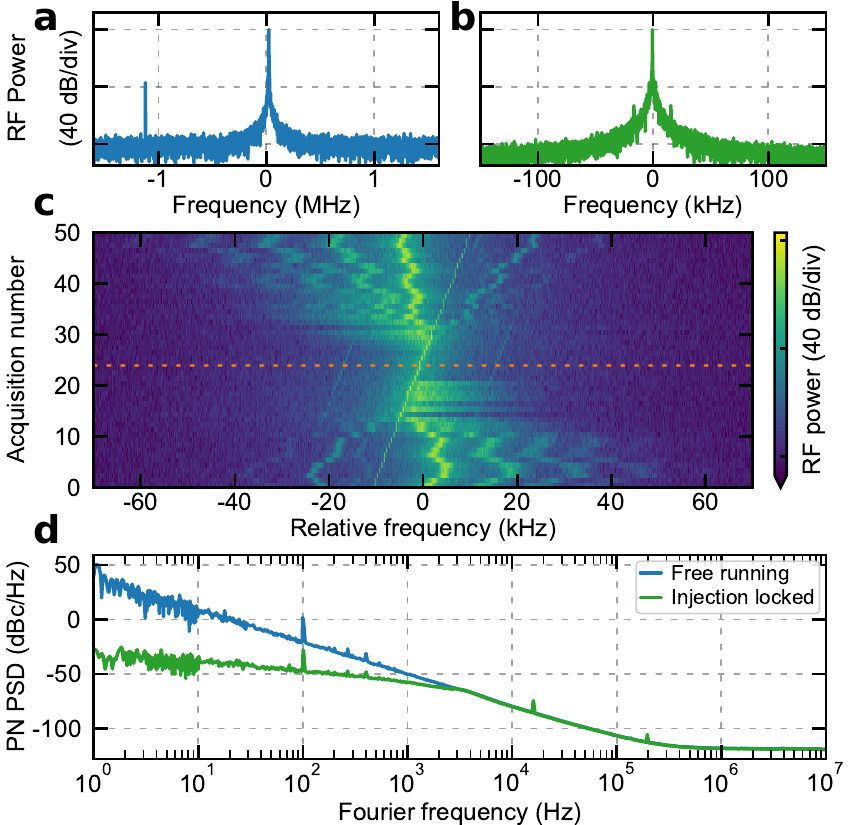}
\caption{Injection locking of the repetition rate by the NIR laser. (a, b) Electrically measured RF spectrum of the QCL with (a) the free running intermode beat and the detection of a weak off-resonant NIR modulation near $-1$~MHz, and (b) the injection-locked intermode beat using 1.5~mW of NIR light with 100\% modulation depth. (c) Stacked RF spectra (50~Hz RBW) with the increasing modulation frequency crossing the natural repetition frequency and causing injection-locking over a range of a few kHz. The dashed horizontal line indexes the acquisition shown in (b). (d) Phase noise power spectral density (PN PSD) of the intermode beat signal in the free-running and injection locked regime.}
\label{fig:inject_lock}
\end{figure}

We thus modulated the EAM of the NIR laser near the repetition frequency near 11.06~GHz while monitoring the generated voltage modulation of the QCL via the dedicated RF extraction port on a RF spectrum analyzer. At low NIR power and close to resonance ($1$~MHz offset), the modulation frequency was picked up by the QCL, which thus acted as a NIR detector (see \autoref{fig:inject_lock}(a)). By slightly increasing the average NIR power with an amplifier to 1.5~mW at the QCL and the RF power on the EAM to 17~dBm (estimated modulation depth close to 100\%), we were able to injection-lock the repetition frequency to the synthesizer and obtain the resolution-limited signal shown in \autoref{fig:inject_lock}(b) at 50~Hz RBW. We obtained a lock range of a few kHz when scanning the modulation frequency across the free running repetition frequency (see \autoref{fig:inject_lock}(c)). The phase noise PSD  was reduced up to a Fourier frequency of about 3~kHz compared to the free-running case (\autoref{fig:inject_lock}(d)). We expect that larger locking ranges could be achieved when using more NIR power, which will be one of the study we will present in another dedicated article.

\section{Discussion}
In this article, we used a low power NIR CW light illuminating the front facet of a MIR QCL frequency comb as an optical actuator for the phase stabilization of a comb line and as a mean to achieve coherent injection locking of the repetition rate. First, by characterizing the response of the QCL, we showed that intensity modulation of the NIR light offers a higher modulation bandwidth compared to conventional drive current modulation. Then, we implemented a stabilization scheme exploiting the NIR light to extend the locking-bandwidth from 300~kHz to over 2~MHz, which resulted in an increase of the SNR by 35~dB and an integrated phase noise as low as 200~mrad. Finally, we showed that the QCL can act as a detector of NIR light modulated at a frequency of 11~GHz, and that it can be injection-locked in such a way. 

We believe that the tighter mutual stabilization enabled by the high bandwidth of the NIR light could lead to higher sensitivities through coherent averaging in DCS~\cite{Chen-natcomm-2018,Komagata-oe-2021}, which is currently one of the main applications of QCL combs. In this regard, a comprehensive comparison with the performance of computational coherent averaging~\cite{sterczewski2019computational} is necessary. The high bandwidth could also allow locking to enhancement cavities for further improvement of the sensitivity~\cite{foltynowicz2013cavityenhanced}.
Moreover, high spectral purity in the MIR is relevant for a variety of applications such as quantum control of molecules~\cite{sinhal2020quantumnondemolition}, tests of fundamental physics~\cite{jansen2014perspective,cournol2019new}, and generally for the study of molecules via high-resolution spectroscopy~\cite{changala2019rovibrational,germann2022methane,agner2022highresolution}.
Thus, we expect our results to facilitate the application of QCL frequency combs in the MIR.

As a further outlook, we anticipate that other wavelengths~\cite{peng2017purified} could be more suitable for orthogonal control without sign reversal of the comb properties such as the offset and repetition frequencies, while perhaps they could also be used to dynamically tune laser parameters such as dispersion, gain, or nonlinearity. Faster modulations from 10~MHz to a few GHz could be investigated as a way to achieve (pure) frequency modulation of the comb. Then, full optical control of the QCL comb merely driven by a battery could be envisaged. To keep the compactness of the device and its low footprint, light could be delivered by fiber or directly generated and modulated on the same chip if low power is sufficient. Optical control could also be investigated in interband cascade lasers~\cite{bagheri2018passively}. Moreover, the recent work on free-space communications using QCL combs~\cite{corrias2022analog} inspires the adaptation of injection locking of the QCL by NIR light for direct conversion of NIR telecommunication streams to MIR signals. We believe that increasing the injected optical power could increase the locking range of the repetition rate to a sufficient level for this application. Further work including theoretical and numerical investigations of the laser dynamics~\cite{villares2015quantum,opacak2019theory} are necessary to understand the full potential of controlling QCLs via optical means.

\begin{acknowledgments}
We thank Stéphane Schilt for support in the early stage of the investigation. We thank Alpes Laser for providing the DFB-QCL used in this work.

We acknowledge fundings from the Schweizerischer Nationalfonds zur Förderung der Wissenschaftlichen Forschung (40B2-1\_176584).

Data underlying the results presented in this paper will be made available on an open server.

\end{acknowledgments}

\bibliography{biblio.bib}

\end{document}